\documentstyle[amssymb,aps,preprint,prb]{revtex}
\draft

\begin{document}
\title{$ $}

\title{Dynamic instabilities and memory effects in vortex matter}

\author{$ $}
\author{Y. Paltiel$^*$, E. Zeldov$^{*,\dagger}$, Y. N. Myasoedov$^*$, H. Shtrikman$^*$, S.
Bhattacharya$^{\ddagger,\S}$, M. J. Higgins$^\ddagger$, Z. L. Xiao$^\P$, E. Y. Andrei$^\P$,
P. L. Gammel$^\dagger$, and D. J. Bishop$^\dagger$}
\address{$ $}
\address{$ $}
\address{$ $}

\address{$^*$Department of Condensed Matter Physics, 
The Weizmann Institute of
Science, Rehovot 76100, Israel}
\address{$^\dagger$Bell Laboratories, Lucent Technologies, Murray Hill, New Jersey 07974, USA}
\address{$^\ddagger$NEC Research Institute, 4 Independence Way, Princeton, New Jersey 08540, USA}
\address{$^\S$Tata Institute of Fundamental Research, Mumbai-400005, India}
\address{$^\P$Department of Physics and Astronomy, Rutgers University, Piscataway,
New Jersey 08855, USA}
\maketitle

\newpage

{\bf Understanding the nature of flow is essential for the resolution of a 
wide class of phenomena in condensed matter physics, ranging from dynamic friction, through pattern formation in sand dunes, to the pinning of charge density waves.
The flux line lattice in type II superconductors serves as a unique model system with tunable dynamic properties. Indeed, recent studies have shown a number of puzzling phenomena including: ({\em i}) low frequency noise 
\cite{marley95,merithew,kwok,danna,maeda}, ({\em ii}) slow voltage oscillations \cite{kwok,gordeev}, ({\em iii}) history dependent dynamic response 
\cite{shobo95,hend96,baner98,baner99,won,degroot}, ({\em iv}) memory of the
direction, amplitude, duration, and even the frequency of the previously
applied current \cite{hend98,xiao}, ({\em v}) high vortex mobility for {\em ac} current with no apparent vortex motion at {\em dc} current \cite{hend98,eva,welp}, and ({\em vi}) strong suppression of an {\em ac} response by small {\em dc} bias \cite{hend98}. Taken together, these phenomena are incompatible with the current understanding of vortex dynamics. By investigating the current distribution across single crystals of 2H-NbSe$_2$ we reveal a generic mechanism that accounts for these observations in terms of a competition between the injection of a disordered vortex phase at the sample edges, and the dynamic annealing of this metastable disorder by the transport current. 
For an {\em ac} current, only narrow regions near the edges are in the disordered phase, while for {\em dc} bias, most of the sample is filled by the pinned disorder, preventing vortex motion. The resulting spatial profile of disorder acts as an {\em active memory} of the previous history.}

In conventional superconductors like NbSe$_2$ the anomalous phenomena are found
\cite{marley95,merithew,shobo95,hend96,baner98,baner99,won,hend98,xiao,eva} in the vicinity of the `peak effect' where the critical current $I_{c}$ increases sharply below the upper critical field $H_{c2}$, as described in Fig. 1a. The peak effect marks a structural transformation \cite{marley95,merithew,shobo95,hend96,baner98,baner99,won,gammel98} of the vortex lattice: Below the peak region an ordered phase (OP) is present, which is dominated by the elastic energy of the lattice and is, therefore, weakly pinned. On approaching the peak region, however, the increased softening of the lattice causes a transition into a disordered vortex phase (DP), which accommodates better to the pinning landscape, resulting in a sharp increase in $I_c$. In high-temperature superconductors like Bi$_2$Sr$_2$CaCu$_2$O$_8$, this situation is equivalent to the second magnetization peak \cite{anis}, where the ordered Bragg-glass phase is believed to transform into a disordered solid \cite{gl,en,vin}.
Figure 1a shows the $I_{c}$ measured at various frequencies. On the high temperature side of the peak effect $I_{c}$ is frequency independent; in this region the DP is thermodynamically stable. In contrast, on the low temperature side, a significant frequency dependence is observed \cite{hend98,eva}; in this region all the unusual vortex response phenomena appear \cite{marley95,merithew,shobo95,hend96,baner98,baner99,won,hend98,xiao,eva}.
As described below, a dynamic coexistence of the OP and a metastable DP is established in this region in the presence of an applied current.
We first outline the proposed mechanism, and then present the experimental evidence. 

The first important ingredient of the proposed model is the observation that in NbSe$_2$ the DP can be readily
supercooled to below the peak effect by field cooling, where it remains metastable, since the thermal fluctuations are negligible \cite{hend96,baner98,baner99,won}. This supercooled DP is pinned more strongly and displays a significantly larger critical current $J_c^{dis}$ as compared to $J_c^{ord}$ of the stable OP. An externally applied current in excess of $J_c^{dis}$ serves as an effective temperature and `anneals' the metastable DP as observed by transport \cite{hend96}, magnetic response \cite{baner99},
decoration \cite{pardo}, and SANS experiments \cite{yaron} on NbSe$_2$.

The second ingredient of the model is the presence of substantial surface barriers \cite{bl}, as observed recently \cite{yossi} in NbSe$_2$. Consider a steady state flow of an OP in the presence of a transport current. In the
standard platelet strip geometry, in a perpendicular field $B$, vortices penetrate from one edge of the sample and exit at the opposite edge. In the absence of a surface barrier, vortex penetration does not require any extra force. As a result, the vortices penetrate close to their proper vortex lattice locations, as dictated by the elastic forces of the lattice. In the presence of a surface barrier, however, a large force is required for vortex penetration and exit, and hence much of the applied current flows at the edges in order to provide the necessary driving force \cite{yossi,dannat,danprl,bur}. The surface barrier is known to be very sensitive to surface imperfections. Therefore, the penetrating vortices are injected predominantly at the weakest points of the barrier, thus destroying the local order and forming a metastable DP near the edge, which drifts into the sample with the flow of the entire lattice. 
(Note that steps on the surface or extended defects in inhomogeneous samples 
could also act as injection points of the DP). The applied current, therefore, has two effects: the current that flows at the edges causes `{\em contamination}' by injecting a DP, while the current that flows in the bulk acts as an {\em annealing} mechanism.  
The observed dynamic instabilities and memory phenomena arise from the fine balance between these two competing processes. 

The annealing process is sensitive to the exact location on the $H-T$ phase diagram. Below the peak effect, the DP is highly unstable and therefore its relaxation time $\tau_{r}$ in the presence of a driving force is very short. As a result, it anneals rapidly over a characteristic `healing' length $L_{r}=v\tau _{r}$, where $v$ is the vortex lattice drift velocity. The corresponding profile of the local critical current $J_c(x)$ should therefore decay from $J_c^{dis}$ to $J_c^{ord}$ over the characteristic length scale $L_r$, as illustrated by the dotted line in Fig. 1b. Note that $L_r$ and $\tau_r$ are generally current dependent and decrease dramatically at elevated currents \cite{hend96}.
On the other hand, near the peak effect the free energies of the DP and
OP are comparable and therefore the `life time' of the disordered phase $\tau _{r}$ and the corresponding $L_{r}$ are very large. 
As a result, the front of the DP, given by $x_{d}(t)$, progressively penetrates into the bulk as shown by the solid line in Fig.
1b, until the entire sample is contaminated. 
{\em In this situation the experimental, steady state dc critical current $I_{c}^{dc}$ does not reflect an equilibrium property, but rather a dynamic coexistence of two phases}. It is given by 
$I_{c}^{dc}=d \int_0^WJ_c(x)dx \simeq dL_{r}J_{c}^{dis}+d(W-L_{r})J_{c}^{ord}$ for $L_{r}< W$, and 
$I_c^{dc}\simeq I_{c}^{dis}=dWJ_{c}^{dis}$ for $L_{r}\gg W$, where $d$ and $W$ are the thickness and width of the sample (neglecting, for simplicity, the surface barrier edge currents, and assuming, for example, an exponential decay of $J_c(x)$ in Fig. 1b).

The {\em ac} response of the system should be distinctly different since the
contamination process occurs only near the edges, where the disordered
lattice periodically penetrates and exits the sample. 
For a square wave $I_{ac}$ of period $T_{ac}=1/f$, 
by the end of the positive half cycle the DP occupies the left edge to a depth of $x_d^{ac}$, as illustrated by the solid curve in Fig. 1b. During the negative half cycle the DP on the left exits the sample, while a DP on the right edge penetrates, until at $t=T_{ac}$ a mirror-image profile is obtained, as shown by the dashed line. 
Assuming $x_d^{ac}<W<L_r$, the effective $I_c$ observed by {\em ac} transport measurement is given by 
$I_c^{ac}\simeq dx_d^{ac}J_c^{dis}+d(W-x_d^{ac})J_c^{ord}$. 
Thus, an {\em ac} current necessarily contaminates the sample less than a {\em dc} current of the same amplitude, and therefore $I_c^{ac}\leq I_c^{dc}$ always, as seen in Fig. 1a. In addition, since $x_d^{ac}$ decreases with frequency, $I_c^{ac}$ should decrease with $f$ explaining the frequency dependence of $I_c^{ac}$ in Fig. 1a. Furthermore, and most importantly, at sufficiently high frequency $I_c^{ac}$ should approach the true $I_c$ of the stable phase. The steep increase of the 881 Hz $I_c^{ac}$ data in Fig. 1a (open circles) therefore indicates that the OP transforms sharply into the DP at the peak effect. In contrast, the smooth behavior of $I_c^{dc}$ reflects rather the dynamic coexistence of the two phases in which $L_{r}$ gradually increases and diverges upon approaching the peak effect from below. From Fig. 1a
we can evaluate $L_{r}$ and $\tau _{r}$. For example, at T=5.1K, $I_c^{dc}\simeq$ 50 mA is about half way between $I_c^{ord} \simeq$ 5 mA and the extrapolated $I_c^{dis}\simeq$ 100 mA, which means that $L_{r} \simeq 0.5W = 170      \mu$m. The $I_c^{dc}$ was measured at a voltage criterion of 1 $\mu$V, which translates into vortex velocity of $v\simeq 4 \times 10^{-3}$ m/sec, and hence $\tau _{r}=L_{r}/v \simeq 4\times 10^{-2}$ sec. This value is well within the range of the relaxation times measured previously \cite{hend96} by applying a current step to the field-cooled matastable DP.

We now provide a direct experimental manifestation of the key aspect of the model, which is the spatial variation of the disorder and $J_c (x)$, and of the transport current distribution that traces this $J_c(x)$ (see Fig. 1b). We have used Hall sensor arrays to measure the {\em ac} transport current self-induced field \cite{yossi,dannat,danprl} $B_{ac}(x)$, which is then directly inverted into the current density distribution $J_{ac}(x)$ using the Biot-Savart law, as described previously \cite{yossi}(see inset to Fig 1a). Figure 2 shows the corresponding current profiles $J_{ac}(x)$ measured at different frequencies. At high $f$, the DP with the enhanced $J_c$ is present only in narrow regions near the edges (481 Hz data). As the frequency is reduced, $x_d^{ac}$ grows and correspondingly the enhanced $J_{ac}(x)$ flows in wider regions near the edges. Note that our measurement procedure (see Fig. 2) provides the time averaged local amplitude of $J_{ac}(x)$, which is much smoother as compared to the sharp instantaneous profiles in Fig. 1b. 

We confirm the above finding independently by measuring the corresponding {\em ac} resistance of the sample $R_{ac}(f)$ as shown in Fig. 3a. At high frequencies most of the sample is in the low pinning OP and therefore $R$ is large. As $f$ is decreased, progressively wider regions near the edges become contaminated with the more strongly pinned DP and thus $R_{ac}(f)$ decreases.
If the applied $I_{ac}$ is larger than $I_{c}^{dc}$, a finite $R$ will be measured at all frequencies, however, if $I_{c}^{ac}\leq I_{ac}\leq I_{c}^{dc}$ (see Fig. 1a) the measured $R$ will vanish as $f\rightarrow 0$, as observed in Fig. 3a. This explains the surprising phenomenon of finite vortex response
to {\em ac} current, while for {\em dc} drive the vortex motion is absent \cite{hend98,eva,welp}. 
From $R_{ac}(f)$ one can directly calculate the width of the disordered regions by noting that $x_d^{ac}$ equals the distance the entire lattice is displaced during half an {\em ac} period, $x_d^{ac}(f)=v/2f=R(f)I_{ac}/2fLB$, where $L$ is the voltage contact separation.
The open circles in Fig. 3b show $x_d^{ac}$ obtained from $R_{ac}(f)$, while the open squares show the $x_d^{ac}$ derived directly from the $J_{ac}(x)$ profiles of Fig. 2. The good correspondence between the two independent evaluations of $x_d^{ac}$ demonstrates the self-consistency of the model. 

Next we address the extreme sensitivity of the {\em ac} response to a small {\em dc} bias \cite{hend98}, as shown in Fig. 4a, where $R_{ac}$ is presented as a function of a superposed $I_{dc}$. A {\em dc} bias of only 10 to 20{\%}
of $I_{ac}$ suppresses $R_{ac}$ by orders of magnitude. This behavior is a
natural consequence of the described mechanism since the {\em dc} bias contaminates the sample very similarly to the pure {\em dc} case, except that $L_r$ is now renormalized as following.  
For $I_{dc}\ll I_{ac}$, the vortices move back and forth during the {\em ac} cycle, with a forward displacement being larger by about $2I_{dc}/I_{ac}$. Therefore, a vortex that enters through the sample edge and reaches a position $x$, accumulates a much longer total displacement path of $xI_{ac}/I_{dc}$. Since the annealing process of the DP depends on the total displacement regardless of the direction, the lattice at this location is thus annealed substantially, as if the effective $L_r$ is reduced to $L_{r}^{eff}\simeq L_{r}I_{dc}/I_{ac}$. Thus, at very small biases, 
the DP is present only within $x_{d}^{ac}$ from the edges, as in
the absence of a bias, where the disordered vortices exit and re-penetrate every
cycle. Vortices that drift deeper into the bulk under the influence
of $I_{dc}$ are practically fully annealed due to the very short $L_{r}^{eff}$. As a result in Fig. 4a the initial decrease of $R_{ac}$ up to $I_{dc}\simeq 2$ mA is relatively small. The corresponding $J_{ac}(x)$ in Fig. 4b at $I_{dc}=1.7 $ mA shows narrow contaminated regions near the edges, very similar to the zero bias case in Fig. 2. However, as $I_{dc}$ is increased, $L_{r}^{eff}$ grows and the bulk of the sample becomes contaminated by the penetrating DP, leading to a dramatic drop of $R_{ac}$. In this situation, $J_{ac}(x)$ at $I_{dc}=5.7$ mA shows a wide region of DP at
the left edge. When $I_{dc}$ is inverted to -5.7 mA, a similar situation 
is observed, but now the vortices and hence the DP penetrate from the right edge, as expected. 

The revealed mechanism readily explains a wide range of additional reported phenomena: ({\em i}) The history of the previously applied current is encoded in the spatial profile of the lattice disorder, which is preserved while the current is switched off due to negligible thermal relaxation. Upon reapplying the current, the vortex system will display a memory of all the parameters of the previously applied current including its direction, duration, amplitude, and frequency, as observed experimentally \cite{hend98,xiao}.
({\em ii}) Application of a current step $I< I_c^{dc}$ to a sample in the OP, results in a transient response which decays to zero since the DP is able to penetrate only a limited distance. The resulting new $I_c$ of the sample is given by the condition that $I_c=I$, as derived by fast transport measurements \cite{xiao,eva}. Such transient phenomena, would also display characteristic times shorter or comparable to the vortex transit time across the sample, in agreement with observations \cite{hend98,xiao}.
({\em iii}) The competition between the contamination and annealing processes is expected to result in local instabilities causing the reported noise enhancement below the peak effect \cite{marley95,merithew} (see also Fig. 4a). 
({\em iv}) Related phenomena should be observed in high-temperature superconductors in the vicinity of the peak effect associated with the melting transition, or near the second magnetization peak, consistent with experiments \cite{kwok,danna,maeda,gordeev,degroot,welp}. 
({\em v}) In high-temperature superconductors there is an additional consideration of thermal activation of vortices over the surface barriers, which may explain the reported slow voltage oscillations \cite{kwok,danna,maeda,gordeev}. If the thermal activation rate is higher or comparable to the driving rate, the slowly injected lattice will be ordered in contrast to the DP injected at higher drives. Thus, at a given applied current, if the bulk of the sample is in the OP, much of the current flows on the edges, rapidly injecting a DP through the surface barrier. Once the bulk gets contaminated, the resulting slower vortex motion causes again injection of an OP. This feedback mechanism can explain the voltage oscillations \cite{kwok,gordeev} in YBa$_{2}$Cu$_{3}$O$_{7}$ and similar narrow band noise \cite{danna,maeda} in YBa$_{2}$Cu$_{3}$O$_{7}$ and Bi$_{2}$Sr$_{2}$CaCu$_{2}$O$_{8}$ 
with characteristic frequencies comparable to the inverse transit time.
({\em vi}) Finally, the described phenomena should be absent in the Corbino disk geometry where vortices do not cross the sample edges. Our studies of NbSe$_{2}$ in this geometry confirm this prediction, as will be published elsewhere.

\newpage 

ACKNOWLEDGEMENTS

We acknowledge helpful discussions with P. B. Littlewood. The work at WIS was supported by the Israel Science Foundation - Center of Excellence Program, 
by the US-Israel Binational Science Foundation (BSF),
and by Alhadeff Research Award. EYA acknowledges support from NSF.

\newpage 

\ FIGURE CAPTIONS

Fig. 1. The experimental setup (inset), the critical current vs. temperature in the vicinity of the peak effect (a), and (b) a schematic plot of the dynamic coexistence of the ordered phase (OP) with a metastable disordered phase (DP).

{\em Experimental}.
Several Fe (200 ppm) doped single crystals of 2H-NbSe$_2$ were investigated. Here we report data on crystal A of $2.6\times 0.34\times 0.05$ mm$^3$ and $T_c=6.0$K, and crystal B of $2.4\times 0.29\times 0.02$ mm$^3$ with 
$T_c=6.05 $K. Four electrical contacts were attached to the top surface for transport measurements, with the voltage contact separation of 0.6 $\pm$ 0.2 mm. The bottom surface of the crystal was attached to an array of 19 2DEG Hall sensors $10\times 10$ $\mu m^2$ each (inset). 
The vortex lattice was initially prepared in the OP by zero-field-cooling to a low temperature at which the DP is unstable, and then slowly heated to the desired temperature in the presence of a constant field $H_a$ applied parallel to the c axis.

(a) The peak effect in critical current $I_c$ in NbSe$_{2}$ crystal A vs. temperature, as measured with a {\em dc}
drive, $I_{c}^{dc}$ ($\square $), and {\em ac} drive, $I_{c}^{ac}$, at 181 
($\blacksquare $) and 881 Hz ($\bigcirc $). The critical current was determined resistively using a voltage criterion of 1 $\mu$V. At low temperatures only the stable OP is present. On the lower temperature side of the peak effect a metastable DP coexists dynamically with the stable OP resulting in a frequency dependent $I_c$. On the high temperature side of the peak effect only the stable DP is present with no anomalous behavior.  

(b) Schematic plot of the local critical current density $J_{c}(x)$ across a crystal of width $W$.
$J_c^{dis}$ and $J_c^{ord}$ are the values of the critical current density in the fully disordered and in the OP, respectively. For $L_{r}\ll W$ the DP relaxes rapidly into OP resulting in the dotted $J_{c}(x)$ in the steady state {\em dc} flow. For $L_{r}>W$ the DP penetrates to a depth $x_{d}(t)$ following the application of {\em dc} current at $t=0$ (solid curve). For an {\em ac} current at $t=T_{ac}/2$ the DP occupies the left edge to a depth of $x_{d}^{ac}$ (solid curve), and symmetrically the right edge at $t=T_{ac}$ (dashed curve).

Fig. 2. Current density profiles $J_{ac}(x)$ in crystal B obtained by inversion of the self-induced field measured by the Hall sensors. Shown are three frequencies $f= 22$ ($\bullet $), 181 ($\square $), and 481 Hz ($\blacksquare $). A square wave {\em ac} current $I_{ac} $ was applied and the corresponding self induced magnetic field $B_{ac}(x)$ across the crystal was measured by the Hall sensors using a lock-in amplifier (see Fig. 1 inset). By using the Biot-Savart law the $B_{ac}(x)$ was directly inverted \cite{yossi} into the current density profiles $J_{ac}(x)$.
The width $x_d^{ac}$ of the highly pinned DP near the edges grows with decreasing frequency as expected.
The measured $J_{ac}(x)$ is the magnitude of the local current density averaged over the {\em ac} cycle period. As a result, $J_{ac}(x)$ reflects a time-averaged superposition of solid and dashed $J_c(x,t)$ profiles in Fig. 1b, which are present separately during the positive and negative half-cycles. Close to the edges the high $J$ is present most of the time, while close to $x_d^{ac}$ it is present only a small fraction of the {\em ac} period as the DP moves in and out of the sample. Therefore, the time-averaged 
$J_{ac}(x)$, decreases smoothly from the edge to $x_d^{ac}$.
A more detailed analysis shows that the first-harmonic measurement by the lock-in amplifier Fourier transforms the sharp instantaneous $J_c(x,t)$ of Fig. 1b into the observed smooth $J_c^{ac}(x)=J_c^{ord}+(J_c^{dis}-J_c^{ord})(1+cos(\pi%
x/x_d^{ac}))/2+(J_c^{dis}-J_c^{ord})(1-cos(\pi (x+x_d^{ac}-W)/x_d^{ac}))/2$. The second and third terms hold at $0\leq x\leq x_d^{ac}$ and $W-x_d^{ac}\leq x\leq W$, respectively, and are zero otherwise.

Fig. 3. Frequency dependence of the resistance $R_{ac}(f)$ and of the width of the disordered regions $x_d^{ac}$ in crystal B.
(a) At high frequencies $R_{ac}$ is large since most of the sample is in the weakly pinned OP. As the frequency is decreased the disordered regions increase and $R_{ac}$ drops sharply, when $x_d^{ac}$ reaches values close to the sample width. In the limit of zero $f$ the resistance is zero since the applied current is lower than $I_c^{dc}$.   
(b) The corresponding width of the DP near the edges 
$x_d^{ac}$ derived from the $R_{ac}(f)$ data ($\bigcirc$) and from the
$J_{ac}(x)$ current profiles of Fig. 2 ($\square $). 
As expected, $x_d^{ac}$ increases monotonically with decreasing frequency.
  The drop of the $x_d^{ac}$ data ($\bigcirc$) at very low frequencies is an artifact. At such frequencies the instantaneous vortex motion is present mainly at the onset of the square wave $I_{ac}$ pulses, and decays towards zero during the pulse \cite{hend98}. In this situation the first-harmonic $R_{ac}$ measurement by the lock-in amplifier underestimates the integrated vortex displacement. In addition to the frequency dependence, $x_d^{ac}$ also changes significantly by varying the amplitude of the {\em ac} current $I_{ac}$. Here $I_{ac}$=10 mA was chosen such that $x_d^{ac}$ becomes comparable to the sample width $W$ at low frequencies. 

Fig. 4. Measured resistance $R_{ac}$ at $I_{ac}=20$ mA as a function of 
{\em dc} bias $I_{dc}$ (a) and the corresponding distribution (b) of the {\em ac} current $J_{ac}(x)$ in crystal A. 
(a) At low bias $I_{dc}\lesssim$ 2 mA only the regions near the edges are contaminated by the DP since $L_r^{eff}$ is very short, resulting in only a moderate decrease of $R_{ac}$. For $I_{dc}\gtrsim$ 2 mA the contamination becomes substantial and the significant decrease of $R_{ac}$ is accompanied by enhanced noise due to the local instabilities during the competing contamination and annealing processes.
At still larger {\em dc} bias most of the sample becomes contaminated and $R_{ac}$ drops below our noise level. The corresponding $J_{ac}(x)$ profiles in (b) show that for positive $I_{dc}=+5.7$ mA ($\bullet $) a substantial part of the sample is contaminated from the left edge where the vortices enter into the crystal, and similarly from the right edge for negative bias $I_{dc}=-5.7$ mA ($\blacksquare $).


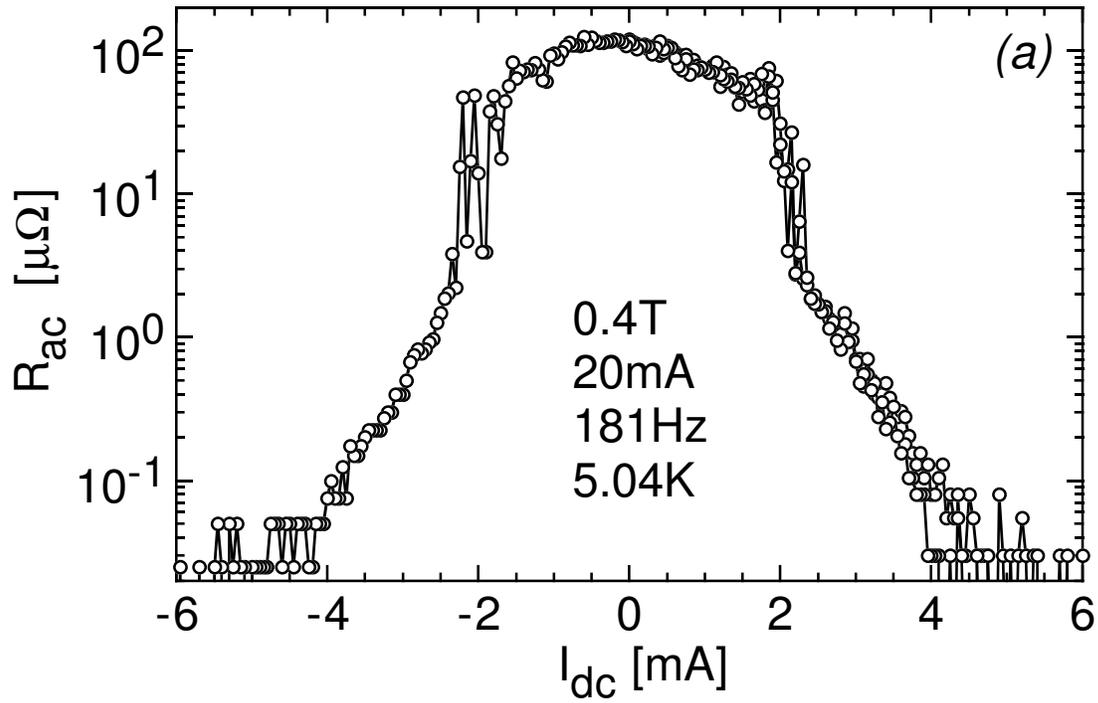
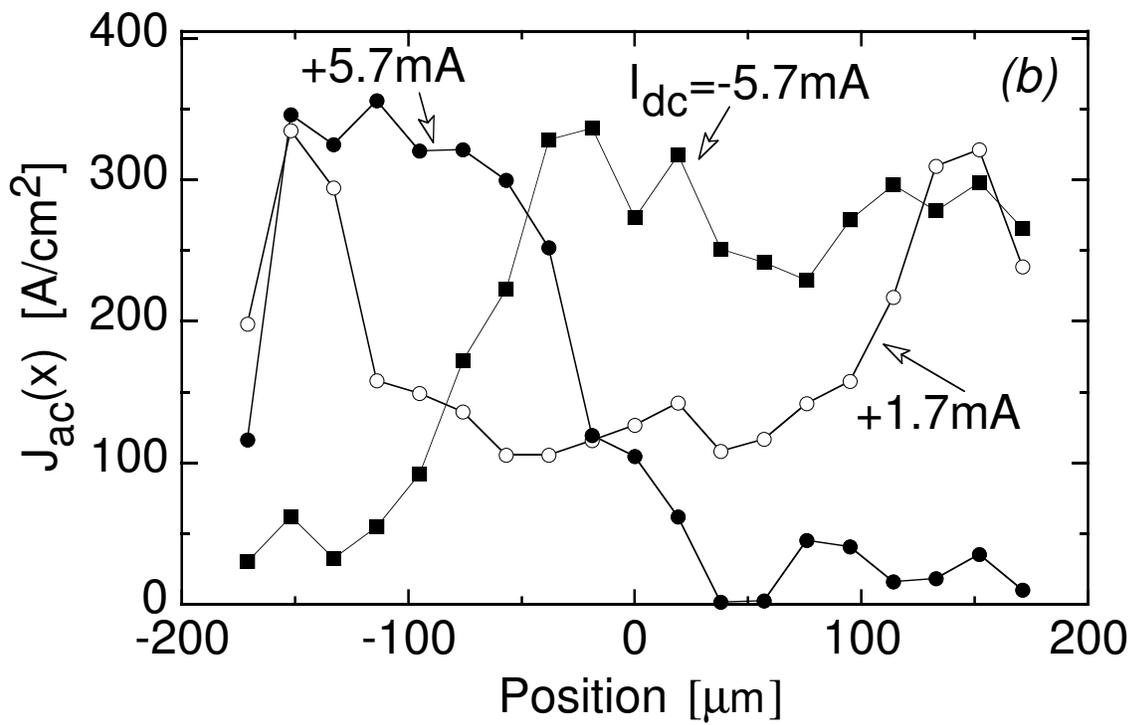
\end{document}